\documentclass[12pt]{JHEP3}
\usepackage{graphicx}
\pdfoutput=1
\newcommand{\be}[1]{ \begin{equation}\label{#1} }
\newcommand{\ee}{\end{equation}}
\newcommand{\bea}[1]{\begin{eqnarray}\label{#1} }
\newcommand{\eea}{\end{eqnarray}}
\newcommand{\eq}[1]{(\ref{#1})}
\def\ZZZ{{\hskip-3pt\hbox{ Z\kern-1.6mm Z}}}
\def\zzz{{\hskip-3pt\hbox{ z\kern-1mm z}}}

\newcommand{\half}{{1\over 2}}

\def\one{{\hbox{ 1\kern-.8mm l}}}
\def\zero{{\hbox{ 0\kern-1.5mm 0}}}

\title{
What is the Simplest Gauge-String Duality?}

\author{
Rajesh Gopakumar$^a$ \\ 
$^a$Harish-Chandra Research Institute, \\
$\;$Chhatnag Road,\\
$\;$Jhusi, Allahabad, India 211019\\
$\;$\email{gopakumr at hri.res.in}}

\abstract{We make a proposal for the string dual to the simplest large $N$ theory, the Gaussian matrix integral in the 'tHooft limit, and how this dual description emerges from double line graphs. This is a specific realisation of the general approach to gauge-string duality which associates worldsheet riemann surfaces to the Feynman-'tHooft  diagrams of a large $N$ gauge theory. We show that a particular version (proposed by Razamat) of this connection, involving integer Strebel differentials, naturally explains the combinatorics of Gaussian matrix correlators. We find that the 
correlators can be explicitly realised as a sum over a special class of holomorphic maps (Belyi maps) from the worldsheet to a  {\it target space} ${\mathbb P}^1$. We are led to identify this target space with the riemann surface associated with the (eigenvalues of the) matrix model. 
In the process, an AdS/CFT like dictionary, for arbitrary correlators of single trace operators, also emerges in which the holomorphic maps play the role of stringy Witten diagrams. Finally, we provide some evidence that the above string dual is the conventional A-model topological  string theory on ${\mathbb P}^1$.}

\preprint{HRI/ST/1101}

\begin{document}

\section{Introductory Remarks}

Even after intensive exploration, the AdS/CFT correspondence has retained its magical and
miraculous flavour, in the manner in which it ties together quantum gauge theories with gravity/closed string theories \cite{Maldacena:1997re, Gubser:1998bc, Witten:1998qj}. We somehow always manage to land on our feet in all  the highly intricate checks of the proposal which have been carried out.  However, the task of physics is to demystify magic and miracles and understand what it is that makes things tick. In that sense we have made very little headway into penetrating into the mystery of the AdS/CFT correspondence.

The original duality between ${\cal N}=4$ Super Yang-Mills and Type IIB closed string theory on
$AdS_5\times S^5$  was motivated by the open-closed string dualities for D-branes in perturbative string theory. However, one believes that this duality (as well as the many related examples) are instances of a much broader and more generic connection between large $N$ gauge theories and perturbative closed string theories. (See \cite{ElShowk:2011ag} for a nice discussion of this connection in a wide context.) This has been, in fact, the motivation behind recent attempts to apply the ideas of AdS/CFT to situations (such as in condensed matter systems) where one does not necessarily have much guidance from string dualities (which mainly give rise to supersymmetric examples anyway). 

\subsection{What is the Simplest Gauge-String Duality?}

Thus from both philosophical as well as practical points of view it is important that we understand the nuts and bolts of the correspondence without relying too much on string dualities. Only then can we hope to map out the space of field theory-string dualities. 
To take apart the mechanism of gauge-string dualities would require us to start with systems that possess a (difficult to achieve) balance between enough structural complexity on the one hand and analytic tractability on the other. Notable examples in this direction are the Kontsevich matrix model 
\cite{Kontsevich:1992ti}
(or equivalently the doubled scaled matrix models) dual to minimal topological string theory 
\cite{Witten:1989ig, Distler:1989ax, Verlinde:1990ku, Gaiotto:2003yb}
as well as Chern-Simons gauge theories in 3d (dual to A-model topological strings on non-compact toric geometries such as the resolved conifold) \cite{Gopakumar:1998ki}. 
These examples are analytically solvable (for all values of the coupling and to all genus in the large $N$ expansion) and therefore afford concrete realisations of the strongest version of the gauge-string correspondence.\footnote{In other words, a duality extending beyond the gravity limit of strong 't Hooft coupling as well as beyond the classical (planar) string limit, to all orders in the $\alpha^{\prime}$ and perturbative ($g_s$) string expansions.}    

However, some of the particularities of these examples have prevented one from drawing too many broad lessons about the general correspondence. One might, at first sight, attribute this to the fact that the examples are primarily topological field/string theories and therefore, have a paucity of physical states (in particular, no Hagedorn density leading to new thermal phases). There is also the apparently serious shortcoming (not unrelated to the previous point) that there is no real Einstein gravity limit in these theories. Therefore various interesting features of the gravitational description such as black holes are absent in the analysis. 

But accepting that we do not learn anything about the puzzles of quantum gravity, the above points, by themselves, do not necessarily invalidate the use of these examples in learning more about the gauge-string correspondence. Rather, these examples could be specially useful for the very reason that they can give an insight into how this correspondence might work away from any gravity limit, in a genuinely stringy regime. Indeed, a gravity limit is quite special in the parameter space and may not even be present in a generic (non-supersymmetric) instance of the correspondence. Thus an over-dependence on the intuition gained from supersymmetric gravity duals might be dangerously misleading. Therefore, topological examples have potentially valuable lessons to teach us about the intrinsically stringy regime of the gauge-string duality. Secondly, topological examples are often embedded in supersymmetric theories as some kind of a closed BPS subsector describing states which preserve a large amount of the supersymmetry. Thus the topological examples implement open-closed string duality in this sector in a transparent way and this may provide a handle on the full duality. 

So what then are the particularities of the examples mentioned (Kontsevich model, Chern-Simons theory) that have obstructed drawing deeper lessons for gauge-string dualities? One of the contributory factors, we believe, is in the {\it nature} of the open-closed string duality in the above examples. One may categorise open-closed string dualities into two broad types which we might call V-type and F-type 
\cite{RGjoburg}. In the V-type dualities, the gauge invariant single-trace operators are associated to {\it vertices} of Feynman graphs which then go over to insertions of closed string operators in the dual worldsheet. In the F-type dualities, the gauge invariant information is associated to {\it faces} of the Feynman graphs which when closed up go over to closed string insertions. See \cite{RGjoburg} (and also \cite{Brown:2010af}) for more on this categorisation and the relation between the two types of dualities.    

In terms of this categorisation  the topological examples above are all of F-type whereas the conventional AdS/CFT duality is of V-type \cite{RGjoburg}. This leads to a different kind of gauge-string dictionary from the usual AdS/CFT one. In particular, the perturbative string states do {\it not} map onto single trace operators. This, and more generally, the relative unfamiliarity with F-type dualities, has led to the above topological 
examples not yet transferring insights directly to conventional AdS/CFT dualities (see however, 
\cite{Berkovits:2007rj, Berkovits:2008qc}). 

The double scaled matrix model \cite{Brezin:1990rb, Douglas:1989ve, Gross:1989vs} which is equivalent to the Kontsevich model in describing minimal topological string theory is, however, of the 
V-type and might seem more promising. 
Unfortunately, in this case, the double scaling limit introduces new technical and physical complications in the bulk-boundary dictionary. This limit is analogous to the BMN limit of ${\cal N}=4$ Yang-Mills which describes the $AdS_5$ pp-wave 
geometry. In some ways it is a bit of a singular limit\footnote{For studying correlation functions as well as the perturbative string expansion, for instance.} to use as a starting point to understand the full theory.  Thus the double scaled matrix model has also not quite integrated itself into the larger body of gauge-string dualities.  

As a combined result of all this we are led to conclude that none of the current "simple" examples of gauge-string dualities have quite achieved the status of a canonical "toy" model for gauge-string dualities.  
In this paper, we will try to make the case that this status should belong to the Hermitian one matrix model in a conventional 'tHooft limit \cite{Brezin:1977sv} (as opposed to a double scaling limit). We will show how, in a perturbative expansion about the 
Gaussian theory, correlation functions of single trace operators go over, in an explicitly describable way, to stringy scattering amplitudes, in a manner appropriate to V-type dualities. We will also see an emergent stringy spacetime from the 0-dimensional matrix model. Many aspects of the 
AdS/CFT bulk-boundary dictionary \cite{Gubser:1998bc, Witten:1998qj} also show up in a skeletal way. The description is, however, intrinsically stringy in this case. In fact, we will find evidence that this dual description is a conventional topological A-model string theory on ${\mathbb P^1}$ 
\cite{Witten:1989ig, Dijkgraaf:1990nc}. All these features makes this example have more in common with the conventional AdS/CFT duality than the earlier topological examples. By not looking at the double scaling limit, one is, in some sense undoing the BMN like limit which led to a somewhat singular description. 

Finally, for us, the most compelling case for taking this model seriously (for uncovering the nuts and bolts of gauge-string duality) is that we arrive at the various conclusions above {\it without invoking any string dualities}. In fact, the approach we will follow is, in essence, the general approach to gauge-string duality advocated in \cite{Gopakumar:2003ns, Gopakumar:2004qb, Gopakumar:2005fx}. So the results of this paper can also be viewed as an explicit  realisation of this approach in a simple context, thus holding promise for generalisation to more complicated cases. More concretely, since the Gaussian matrix model also describes the sector of $\half$-BPS Wilson loops in ${\cal N}=4$ Super Yang-Mills \cite{Pestun:2007rz}, the results here should directly apply to this sector of the dual string theory on $AdS_5\times S^5$. 

\subsection{Outline of the Paper}

Since several different ingredients, all of which may not be entirely familiar to the reader, go into the conclusions of the paper, here we will give an extended outline of the thread of argument. This can serve as a roadmap for the different sections of the paper. 

The Hermitian one matrix model has played a very important role in many areas of theoretical physics. 
\be{hermit}
Z[t_k] = \int [dM]_{N\times N} e^{-N {\rm Tr} M^2 +\sum_k t_k N {\rm Tr} M^{2k}}.
\ee
We will view this expression as the generating function of  arbitrary $n$-point correlators 
$\langle \prod_i^n {\rm Tr} M^{2k_i} \rangle$ in the Gaussian matrix model. 
This is the analogue of evaluating gauge invariant correlators in a free large $N$ gauge theory. 
The main focus here will be on describing the string dual to these correlators. (One could include odd powers in addition, though we will find that the above case is nicer in some ways.)
 
The double scaling limit of this model, $N\rightarrow \infty ,  t_k\rightarrow t_k^{(0)}$ keeping certain combinations of $N, (t_k-t_k^{(0)})$ fixed played an important role in the study of simple, solvable string backgrounds \cite{Brezin:1990rb, Douglas:1989ve, Gross:1989vs}. As we mentioned above, this limit appears to be a bit singular from the point of view of gauge-string duality. 
We will be focussing on the conventional 'tHooft limit of this model ($N$ large but finite,  $t_k$ fixed), which has been relatively less explored from the point of view of gauge-string duality 
\cite{Razamat:2008zr, Razamat:2009mc}. 

Our starting point will be an observation by de Mello Koch and Ramgoolam \cite{Koch:2010zza} 
on $n$-point trace correlators in the Gaussian model.\footnote{Closely related observations were made by Itzykson and collaborators in the early nineties \cite{Itzykson:1990zb, DiFrancesco:1992cn, itzbauer}.} 
As we will review in Sec.2, they show that the combinatorics of the contributions to such correlators from the various Wick contractions has a natural interpretation as a sum over contributions from a special class of holomorphic maps (called "clean" Belyi maps by mathematicians). These are maps from genus $g$ worldsheets to 
a target space which is a 
${\mathbb P}^1$. The special nature of the maps is that they are covering maps with only {\it three} branch points (at say, $(0,1,\infty)$) with simple ramification over (the inverse images of) one of the points. This interpretation of the expression for correlators is very much like that of Gross and Taylor \cite{Gross:1992tu, Gross:1993hu, Gross:1993yt, Cordes:1994sd} for the exact answer for the partition function and Wilson loops of $2d$-Yang-Mills theory. The interesting feature of Belyi maps is that they do not exist for an arbitrary point on the moduli space of riemann surfaces. In fact, the riemann surfaces that do admit Belyi maps are the special set of {\it Arithmetic} riemann surfaces. Namely, those defined by polynomial equations over the field of {\it algebraic numbers} $\bar{\mathbb Q}$ (rather than over ${\mathbb C}$). These comprise a discrete set of points on the moduli space of riemann surfaces. Thus the combinatorics suggests that the correlators get contributions only from worldsheets which are arithmetic \cite{Koch:2010zza}.

We will show that this suggestive form of the combinatorics is indeed realised, in a very explicit way, in the general approach to gauge-string duality advocated in 
\cite{Gopakumar:2003ns, Gopakumar:2004qb, Gopakumar:2005fx}. In this 
approach one starts with the perturbative Feynman-'tHooft diagrams contributing to an $n$-point correlator. As we will briefly review in Sec. 3, by organising these diagrams in a natural topological way and writing them in a proper time representation, one can make a correspondence to the riemann surfaces which arise by "gluing up" the ribbon Feynman graphs (in a so-called Strebel worldsheet gauge). Summing over the different graphs and integrating over the proper times basically covers the moduli space of inequivalent riemann surfaces in a unique way. A precise identification was, in fact, proposed between the proper times and the parameters of the moduli space (in the Strebel parametrisation)
\cite{Gopakumar:2005fx}.  This enables one, in principle, to directly convert the gauge theory expressions (of genus $g$ in an $n$-point function) into integrals over the moduli space ${\cal M}_{g,n}$. The latter would then be expressions for the $n$-point scattering amplitude in the string dual. From a knowledge of the integrand we expect to be able to reconstruct the string dual. See \cite{Furuuchi:2005qm,  Aharony:2006th, David:2006qc, 
Yaakov:2006ce, Aharony:2007fs, David:2008iz, Brown:2010pb} for further developments along these lines.

In the particular case of the Gaussian matrix model in 0-dimensions (where there 
is no real proper time) Razamat has proposed a modification of the original proposal \cite{Razamat:2008zr, Razamat:2009mc}. 
He identifies (instead of the proper time) the {\it number} of (homotopic) Wick contractions, 
between any two vertices, with the corresponding strebel length parameter. Thus the strebel length 
takes only non-negative integral values. This can also be motivated from the triangulation picture of riemann surfaces where the integrality of these lengths is the conventional discreteness of the triangulation (or generally "polygonulation"). However, it is more accurate to view the riemann surfaces (corresponding to integer Strebel parameter) as special points in the moduli space of riemann 
surfaces. Remarkably, by a result of Mulase and Penkava \cite{mulpenk1}, 
these are the very points which admit Belyi maps, which as we mentioned, correspond to arithmetic riemann surfaces.  Thus, we seem to arrive from two very different points of view, at the conclusion that the dual string amplitudes get localised contributions from the same special set of arithmetic points on the moduli space of riemann surfaces. 

We will thus, in Sec.4, find that the worldsheet 
associated (by the above construction) to each Feynman ribbon graph admits Belyi maps.
These can, in fact, be constructed in a fairly explicit way by gluing together the expressions 
for each strip (Feynman ribbon edge). We find that the contribution to a {\it planar} 
$n$-point correlator is 
given by a sum over {\it general} Belyi maps (i.e. without any restriction to "clean" Belyi maps), of fixed degree, to a target space ${\mathbb P}^1$. We will also see that a natural two fold cover of this 
${\mathbb P}^1$ makes its appearance. This will account for the fact (at the planar level) that the maps that are suggested by the combinatorics of \cite{Koch:2010zza} are of the special kind called "clean" Belyi maps (with simple ramification indices at one of the points). 

We next turn to the physical interpretation of these constructions in Sec. 5. The double cover 
${\mathbb P}^1$ that appears here has a simple interpretation. It is precisely the riemann surface associated with the eigenvalues of the Gaussian matrix model which has a single cut. It has a canonical differential
\be{abdiff }
\omega \propto {dX \over Y} \qquad Y=\sqrt{X(1-X)}.
\ee
This is not to be confused with the {\it worldsheet} riemann surface which is constructed from the Feynman-'tHooft ribbon graphs. The two end points of the cut (which are at $0$ and $1$ in our conventions) are exactly (two of) the points where the covering map are  branched. The third branch point is at $\infty$ in the $X$ plane. The explicit Belyi maps show that this is precisely where the operators are inserted. 

This is as we would expect in AdS/CFT where the correlators correspond to insertions at the boundary of the dual spacetime (alternatively, the UV  of the field theory). Our target space seems to behave like the complex plane with a one dimensional boundary which is identified into the point at 
$\infty$. This fits in 
with the fact the "UV of the matrix model" is zero dimensional (a point). 
Nevertheless, there is an emergent spacetime which arises from the repulsion of the eigenvalues and is reflected in the presence of the cut in the interior of the $X$-plane (the "IR of the matrix model"). As we will see, this is naturally interpreted as the spatial part of the (holomorphic) spacetime seen by the closed strings. The branching corresponds to the splitting and joining of these strings which seems to occur only at the endpoints of the space. 

In Sec.6 we propose a connection of the above stringy picture, which arises directly from the matrix model through the general constructions of \cite{Gopakumar:2005fx, Razamat:2008zr, Razamat:2009mc}, with more conventional worldsheet theories. We attempt to spell out a relation between the planar connected correlators in the Gaussian matrix model with genus zero correlation functions in the A-model topological string theory on 
${\mathbb P}^1$. 
\be{gausstopcorr}
\langle {\rm Tr} M^{2k_1}\ldots  {\rm Tr}M^{2k_n}\rangle_c \sim \langle  \sigma_{2k_1-1}(Q)\sigma_{2k_2-1}(Q)\sigma_{2k_3}(Q)\ldots 
\sigma_{2k_n}(Q)\rangle_0
\ee
Here $\sigma_n(Q)$ are the gravitational descendants of the 
Kahler class $Q$ of the ${\mathbb P}^1$.
As we will see, such a relation reproduces an important feature of the worldsheet construction of Sec.4, namely, that the Belyi maps are of fixed degree $k=\sum_ik_i$. This is precisely what the RHS correlator gives, thanks to a selection rule.
We then explicitly compute the answers for the simplest set of planar connected correlators in the Gaussian model $\langle {\rm Tr} M^{2k}\rangle_c$ and $\langle {\rm Tr} M^{2k_1}{\rm Tr} M^{2k_2}\rangle_c$. 
We use conventional recursion relations to compute the corresponding correlators in the topological string theory and find they match.

Sec.7 concludes with a  summary and discussion of the different aspects that need to be clarified in this construction of the string dual. Two appendices contain details of computations relevant to Sec. 6. 

\section{Gaussian Correlators as a Sum over Belyi Maps}

It was observed in \cite{Itzykson:1990zb, DiFrancesco:1992cn, Koch:2010zza} that $n$-point correlators in the Gaussian model
\be{gcorr}
\langle \prod_i^n {\rm Tr} M^{2k_i} \rangle = \int [dM]_{N\times N} e^{-N {\rm Tr} M^2} \prod_i^n {\rm Tr} M^{2k_i}
\ee
can be simply expressed in terms of particular sums over permutations as follows.
The basic Wick contraction is given by
\be{wick}
\langle M^i_jM^k_l \rangle = N^{-1}\delta^i_l\delta^k_j
\ee
which implies that for $2k$ matrix elements we have a sum over all disjoint pairs of contractions 
\be{wick2}
\langle M^{i_1}_{j_1}M^{i_2}_{j_2} \ldots M^{i_{2k}}_{j_{2k}} \rangle = N^{-k}\sum_{\alpha\in [2^k]}\delta^{i_1}_{j_{\alpha(1)}}\delta^{i_2}_{j_{\alpha(2)}}\ldots \delta^{i_{2k}}_{j_{\alpha(2k)}}.
\ee
Here $\alpha$ is a permutation in $S_{2k}$ in the conjugacy class labelled by $[2^k]$ i.e. purely by transpositions which label the pairs being contracted. 
We are actually interested in the correlator in \eq{gcorr} in which the matrix indices of the $2k=2\sum_i k_i$ matrix entries are traced together in a specified way. This can be represented by another permutation  $\beta \in S_{2k}$ which is in the conjugacy class specified by the cycle structure
$(2k_1)(2k_2)\ldots (2k_n)$ with each cycle corresponding to the corresponding trace in the correlator. 
Combining this with \eq{wick2}, we can evaluate the correlator 
\be{correv}
\langle \prod_i^n {\rm Tr} M^{2k_i} \rangle = \sum_{\alpha\in [2^k]}N^{C_{\gamma}-k}.
\ee
Here $\gamma \in S_{2k}$ is the (inverse of the) permutation given by the composition of the permutations
$\alpha$ and $\beta$ i.e.  $(\gamma^{-1} = \alpha\cdot \beta)$. It is clear that the cyclic structure of 
$\gamma$ is what controls the matrix index contractions once the Wick contractions are performed. In fact, the additional power of $N$ (which counts the completed cycles amongst the indices) is given by the number of disjoint cycles $C_{\gamma}$ in the permutation $\gamma$. 

Thus the three permutations $\alpha ,\beta , \gamma$ (with $\alpha\cdot\beta\cdot \gamma=1$) determine the correlators. We can give a simple interpretation for them. They are associated respectively with the edges, vertices and faces of the Feynman-'tHooft graph contributing to a given Wick contraction.  
There are twice $2k$ half edges in total from all the vertices. The permutation $\alpha$ pairs off these half edges by the Wick contraction. $\beta$ is the (specified) cyclic structure of the correlators and thus associated with the vertices. Finally, since $\gamma$ counts the power of $N$ it is easy to see that it must be associated with the number of distinct colour loops i.e. number of faces. 

For the normalised correlator $N^{n}\langle \prod_i^n {\rm Tr} M^{2k_i} \rangle$, the answer can be written in the suggestive form \cite{Koch:2010zza} 
\bea{corrfin}
N^{n}\langle \prod_i^n {\rm Tr} M^{2k_i} \rangle &=& 
\sum_{\alpha, \gamma}\delta(\alpha\cdot\beta\cdot \gamma)N^{C_{\gamma}+C_{\beta}-k} \cr
&=& 
\sum_{\alpha, \gamma}\delta(\alpha\cdot\beta\cdot \gamma)N^{2-2g}.
\eea 
where 
\be{rhform}
2-2g= 2k(2- 2\times 0) -B
\ee
with $B=(2k-C_{\alpha})+(2k-C_{\beta})+(2k-C_{\gamma})$.
(Note that $C_{\alpha}=k$)
Here \eq{rhform} is the Riemann-Hurwitz formula describing covering maps of degree $2k$ from a genus $g$ riemann surface to a target space (in this case of genus $0$) and $B$ is the branching number. 

Branched coverings of a riemann surface can be characterised by conjugacy classes of the permutation group. Each such conjugacy class labels the so-called ramification 
profile of a branchpoint. Basically, the different cycles of the conjugacy class 
describe how the different covering sheets are 
permuted about the branchpoint. The branching number at each such point is given by the sum of the branchings for each of the cycles and thus equal to $(2k-C)$ where $C$ is the number of cycles in the permutation. The above expressions suggest therefore that we have branched covers of degree $2k$ 
which are branched over three points on the target sphere. The three branchpoints are labelled by 
the permutations $\alpha ,\beta , \gamma \in S_{2k}$. The constraint $\alpha\cdot\beta\cdot \gamma=1$
reflects the fact that the target space is of genus zero and hence a closed loop around the three branchpoints should be trivial and thus correspond to the identity permutation on the sheets. 
Note that $\alpha\in [2^k]$  implies a ramification profile, about the inverse images of one of the points, which is  {\it simple} (which means that it is a sum over double covers since all cycles are of length two).   

We should note that we have been considering all contributions to the $n$-point correlators which includes the disconnected graphs as well. This means that there will be corresponding disconnected worldsheets in the sum. In such cases we should really view the Riemann-Hurwitz formula 
as
\be{rhform2}
2-2g_{eff}= 2k(2- 2\times 0) -B
\ee
where $(2-2g_{eff})=\sum_l(2-2g_l)$ with each of the disconnected components, labelled by $l$, having genus $g_l$. We should really then view the connected contribution as coming from the usual logarithm of the full generating function $Z[t_k]$. This will have an expansion in terms of connected surfaces, each of genus $g$. From now on, we will assume this has been done and when we speak of correlators, we will refer to the connected pieces and their corresponding worldsheets. The matrix correlators therefore generate the so-called Hurwitz numbers which count holomorphic branched covers to ${\mathbb P}^1$.
Note that Hurwitz numbers have recently been discussed in the context of gauge-string duality for 
$AdS_3$ \cite{Pakman:2009ab}.

Thus the expression for the contribution to the (connected) correlator at genus $g$ in the large $N$ expansion 
has the suggestive interpretation as a sum over holomorphic maps from a worldsheet $\Sigma_g$ to a target 
${\mathbb P}^1$ which are branched exactly over three points with the ramification indices over one of the points being simple. Such maps are what mathematicians call "clean" Belyi maps. The adjective "clean" refers to the fact that the ramification indices on the inverse images of one of the points is simple. For a general Belyi map of degree $2k$ there are no restrictions on $\alpha, \beta , \gamma \in S_{2k}$ except for the relation 
$\alpha\cdot\beta\cdot \gamma=1$. 

It is a famous theorem of Belyi that a riemann surface admits a Belyi map if and only if it is {\it arithmetic}. In other words the defining equation for the riemann surface should be expressible in the field of algebraic numbers $\bar{\mathbb Q}$. These form a discrete (but dense) subset of the moduli space of riemann surfaces ${\cal M}_{g,n}$. Thus Belyi maps are of number theoretic significance and play a central role in Grothendieck's programme of {\it Dessins d'Enfants} to understand the absolute Galois group. Some of these aspects are reviewed and studied from a physics perspective in \cite{Ashok:2006du, Ashok:2006br, Jejjala:2010vb}.   

For us, at the moment, the important lesson to take away is that while the combinatorics of
correlators in the Gaussian model seem to admit a natural stringy interpretation, the contributions seem to be coming from discrete points in ${\cal M}_{g,n}$. At first sight it is difficult to see how a continuum string amplitude might have discrete support on ${\cal M}_{g,n}$. Later we will give examples to illustrate the genericity of this phenomenon for topological string theories. We will also see how the sum over maps of a fixed degree can also be naturally realised.

\section{Worldsheets from Feynman Graphs}

A general answer to the question of precisely how large $N$ field theories reorganise themselves into string theories has been proposed in \cite{Gopakumar:2003ns, Gopakumar:2004qb, Gopakumar:2005fx}. Essentially the observation is that there is a natural way in which perturbative Feynman graphs can be glued up into inequivalent (punctured) riemann surfaces of the appropriate large $N$ genus. We will quickly review this construction below (see \cite{Gopakumar:2005fx, Aharony:2006th} for a more complete picture). Viewing the Feynman diagrams in a proper time representation associates a natural length parameter  to the edges of the graph (and therefore also to the dual graph). 

The proposal of \cite{Gopakumar:2005fx} equates the (inverse) of the proper time with the strebel length which parametrises the moduli of riemann surfaces. Integrating over the proper times (and summing over inequivalent graphs) then covers the moduli space ${\cal M}_{g,n}$
(more exactly ${\cal M}_{g,n}\times {\mathbb R}_{+}^n$) once. This is what one expects for a dual string amplitude for the $n$-point function of gauge invariant operators. 

\begin{center}
\vskip 2pt
\resizebox{250pt}{200pt}{\includegraphics{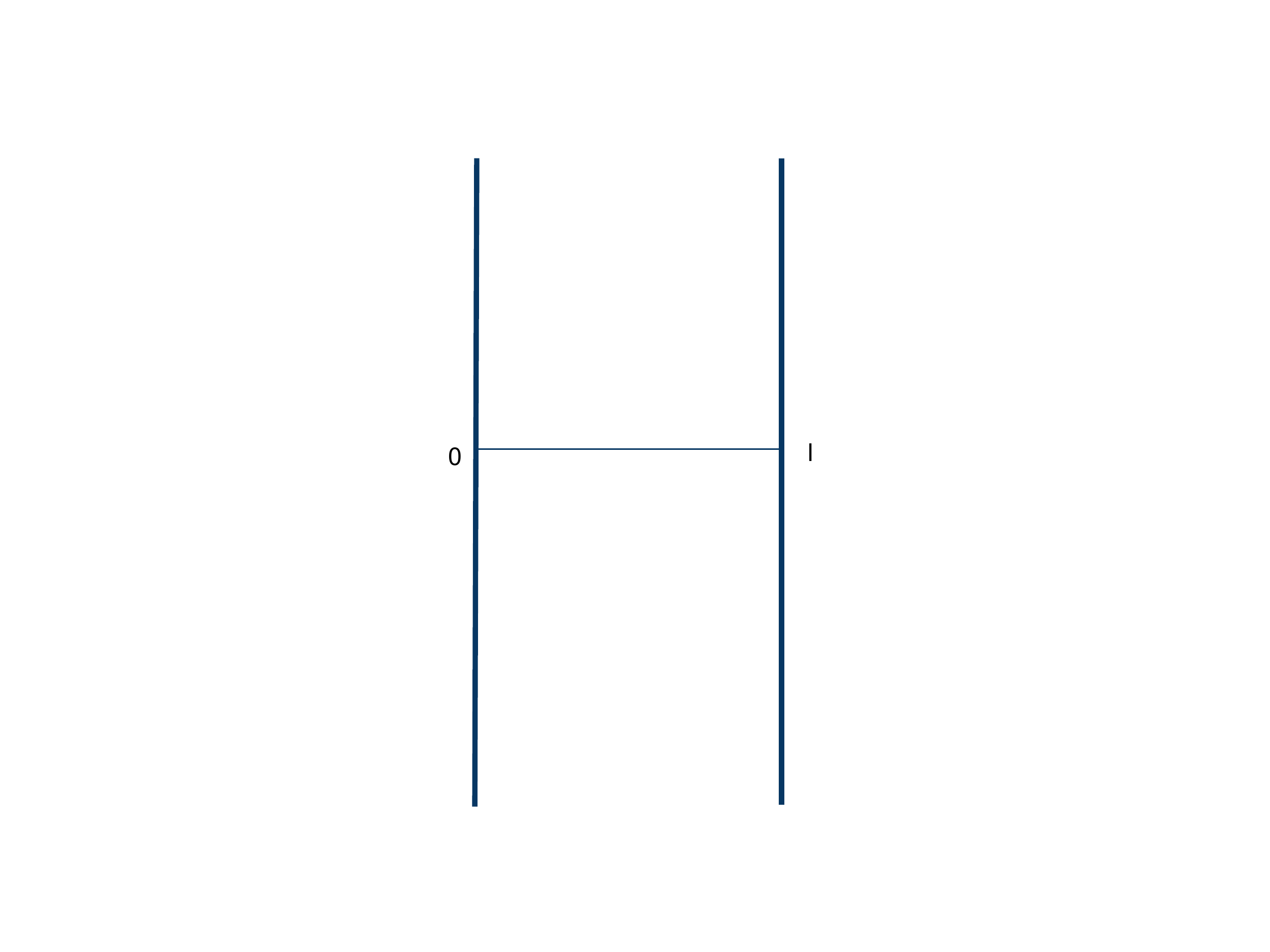}}
\end{center}
\begin{center}
Fig.1: Feynman Ribbon Edge As A Vertical Strip of Width $l$.
\end{center}

\subsection{Gluing Together Ribbon Graphs}

Let us describe  how a riemann surface can be associated to a Feynman ribbon graph 
\cite{Gopakumar:2005fx} (with a length $l_i$ associated to each edge equal to the inverse proper time $\sigma_i={1\over \tau_i}$.).  This follows the general construction of \cite{mulpenk1} which associates to the critical graphs of a Strebel differential a riemann surface. To the $i$th edge of the Feynman graph we associate a vertical 
strip in the complex plane of width $l_i$ i.e. the region $(z \in {\mathbb C} | 0 \leq {\rm Re} z \leq l_i)$. 
On this strip we 
have the differential $dz^2$. We will associate the (horizontal) interval $(0 \leq z \leq l_i)$ with the corresponding dual edge of the dual graph to the original one. $ l_i$ is the strebel length associated to this strip, or equivalently, the dual edge.

We now need to glue together all the different strips (Feynman propagators) to form a worldsheet. Consider a face of the Feynman graph which 
would be (generically) triangular i.e. bounded by three edges. Equivalently, this is a three fold vertex in the dual graph.  
We patch the corresponding three 
strips together  
in a way familiar from open string field theory.  We use a new 
coordinate 
\be{facemap}
w\propto z^{2\over 3}.
\ee 
This maps the part of each strip in the
vicinity of the 
vertex into a wedge of angle ${2\pi \over 3}$. The three wedges 
are glued together in the $w$-plane (see Fig.3). 

\begin{center}
\vskip 2pt
\resizebox{160pt}{120pt}{\includegraphics{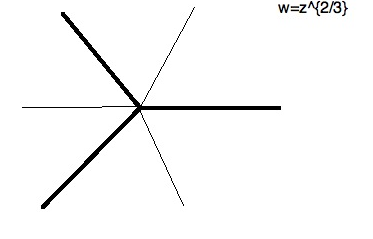}}
\end{center}
\begin{center}
Fig. 2: Gluing Three Strips Together at A Dual Vertex (Triangular Face).
\end{center}

The individual differentials $dz^2$
for each strip are now transformed into a differential of the form 
$wdw^2$ in the $w$-plane which smoothly overlaps between 
the different strips in the vicinity of a vertex. This is the process of gluing up of holes which we expect in open-closed string duality. 
For a $m$-gonal face of the Feynman graph (i.e. a $m$-fold vertex of the dual graph), 
we use a $z^{2\over m}$ map for gluing
the strips together. In particular, note that for the case $m=2$, i.e. when we have two edges 
between the same pair of vertices, the gluing is trivial ($w=z$) and we just place the strips next to each other to get a strip of combined width $l_1+l_2$. 

Similarly, when we have several Feynman edges meeting at a vertex (operator insertion), we glue
halves of the corresponding strips (e.g. ${\rm Im} z \geq 0$) so that they form a semi-infinite cylinder.  
This can be done  
by an exponential map 
\be{vertmap}
u(z)\propto \exp{{2\pi iz\over p_a}}.
\ee
Here $p_a=\sum_i l_{ia}$ where $l_{ia}$ are the widths associated to the various edges incident on the vertex labelled by $a$. This map transforms the upper halves of the various strips into pizza slice shaped regions which are glued together into a disc  (see Fig.3). Note that $z=i\infty$ is mapped into the origin 
$u=0$ which is the vertex where all the strips meet.  
Note also that the differential $dz^2$ in the individual strips 
transforms into a differential of the form 
$$\phi(u)du^2=-{p_a^2\over (2\pi)^2}{du^2\over u^2}.$$
This exhibits the characteristic double pole at $u=0$ with residue $p_a$.

\begin{center}
\vskip 2pt
\resizebox{160pt}{120pt}{\includegraphics{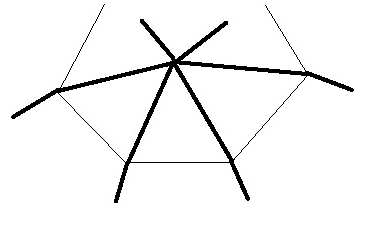}}
\end{center}
\begin{center}
Fig.3: Gluing Together Strips at a Dual Face ($k$-fold Vertex)
\end{center}

These gluing rules of the strips specify a complex structure and define a riemann surface of genus $g$
with $n$ punctures \cite{mulpenk1}. 
The specification of the strebel lengths $l$ picks out a point in the extended moduli space ${\cal M}_{g,n}\times {\mathbb R}_{+}^n$ where the ${\mathbb R}_{+}^n$ piece describes the residues $p_a$ at each of the $n$ vertices. It follows from a theorem of Strebel that as we let $l_i$
take all positive values, we cover ${\cal M}_{g,n}\times {\mathbb R}_{+}^n$ exactly once.

\subsection{Razamat's Modification}

In the case of a 0-dimensional matrix model, we do not have any natural notion of a proper time to associate to edges of the graph. Razamat has proposed an interesting modification of the above proposal in this case. His idea is that the length associated to a given contracted edge should be exactly {\it one}. This implies that the widths of the strips in the above construction are always one. Note, however, that when we have multiple contractions between the same pair of vertices, then the effective width of the glued up parallel strips is equal to the number of (homotopically equivalent) wick contractions.  Thus the strebel lengths associated to a given Feynman diagram are always positive integers. Equivalently the {\it lengths of the edges of the dual graph} are taken to be equal to one (or integral multiples thereof if there are multiple parallel wick contractions in the original graph). 

This assignment, which appears unusual at first sight,  is actually the same as in the discrete dynamical triangulation approach to matrix models. In this approach the {\it dual} graph to the matrix model ribbon graphs is taken to be a discrete triangulation (polygonulation) of a riemann surface \cite{David:1984tx, Kazakov:1985ds, Ambjorn:1985az}. The riemann surface is pieced together with discrete {\it equilateral} triangles  (or generally, regular polygons) and the curvature is assumed to be concentrated at the (dual) vertices (See Fig. 4). We can directly identify the integer lengths of the sides of the polygons with the (integer) strebel lengths. 

But the interpretation here is different. We {\it do not} view the riemann surface as some discrete approximation to a continuuum surface. Instead we view the integer strebel lengths as picking a special set of points in the moduli space ${\cal M}_{g,n}\times {\mathbb R}_{+}^n$. The riemann surface, as reviewed above, consists of 
piecewise cylinders glued together as at any generic point in the moduli space. 

\begin{center}
\vskip 2pt
\resizebox{250pt}{180pt}{\includegraphics{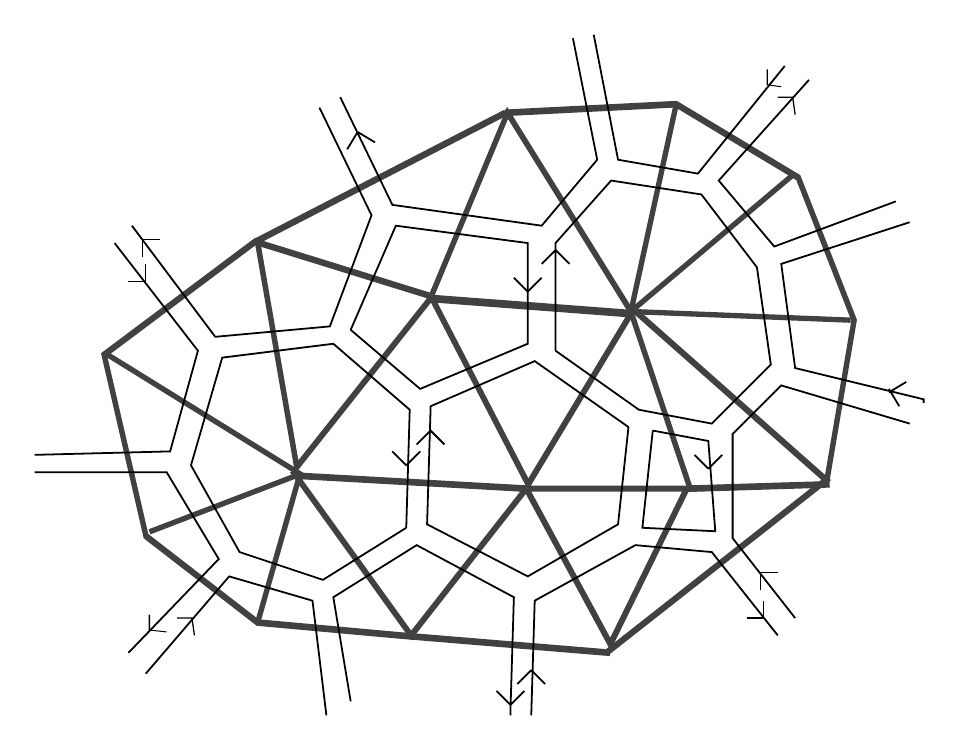}}
\end{center}
\begin{center}
Fig. 4: A Piece of A Triangulation Dual to A Matrix Model Ribbon Graph.
\end{center}

Integer strebel lengths have also been considered earlier as some kind of discretisation of moduli space \cite{Ambjorn:1992gw, Chekhov:1993ee}. It was argued by \cite{Chekhov:1993ee} that this discretisation continues to capture topological information of ${\cal M}_{g,n}$ such as intersection numbers (see also the recent work \cite{Mulase:2010gw}).  In fact, the authors \cite{Ambjorn:1992gw, Chekhov:1993ee} construct a matrix model, the so-called Kontsevich-Penner model, which captures this discretisation. It turns out that open-closed-open string duality \cite{RGjoburg} precisely relates this matrix model to the Gaussian matrix model. 

Remarkably enough, these special points with integer strebel length correspond precisely to those points in the moduli space which admit Belyi maps \cite{mulpenk1}. We will explicitly construct the (local) Belyi maps for the surface with integer strebel lengths, in the  next section. 

\section{Explicit Belyi Maps}

We will now write down (adapting a construction of \cite{mulpenk1}) the explicit maps to 
${\mathbb P}^1$ when the strebel lengths are positive integers.  We do this by locally 
specifying the map from a strip of width {\it one} (corresponding to a single Wick contraction). 
When combined with the gluing procedure for putting the strips together into an $n$-punctured riemann surface of genus $g$, we will see that the maps are branched over precisely three points. For planar graphs these will turn out to be {\it arbitrary Belyi maps of degree $k$}. To connect with the clean Belyi maps (of degree $2k$) of Sec.2, we will see that we need a double cover of the above ${\mathbb P}^1$. This double cover will also arise naturally from our construction. 

For the moment let us restrict ourselves to planar graphs contributing to a (connected) 
correlator of the form
$\langle \prod_i^n {\rm Tr} M^{2k_i} \rangle_c$. We will comment on the case of general genus later. 
For a strip of width one i.e. the region $(z \in {\mathbb C} | 0 \leq {\rm Re} z \leq 1)$, the map to ${\mathbb P}^1$ which covers it {\it exactly once} is
\be{belstrip}
X(z)=\sin^2{\pi z\over 2}. 
\ee
We see that this maps $[0,1]$ to $[0,1]$ and $\infty$ to $\infty$. The vertical edge of the strip at $z=0$
are mapped to the negative real axis in the $X$ plane. While the edge at $z=1$ is mapped to the interval 
$[1,\infty)$. We notice that this leads to a cut between $0$ and $1$ in the target space. 

\begin{center}
\vskip 2pt
\resizebox{240pt}{180pt}{\includegraphics{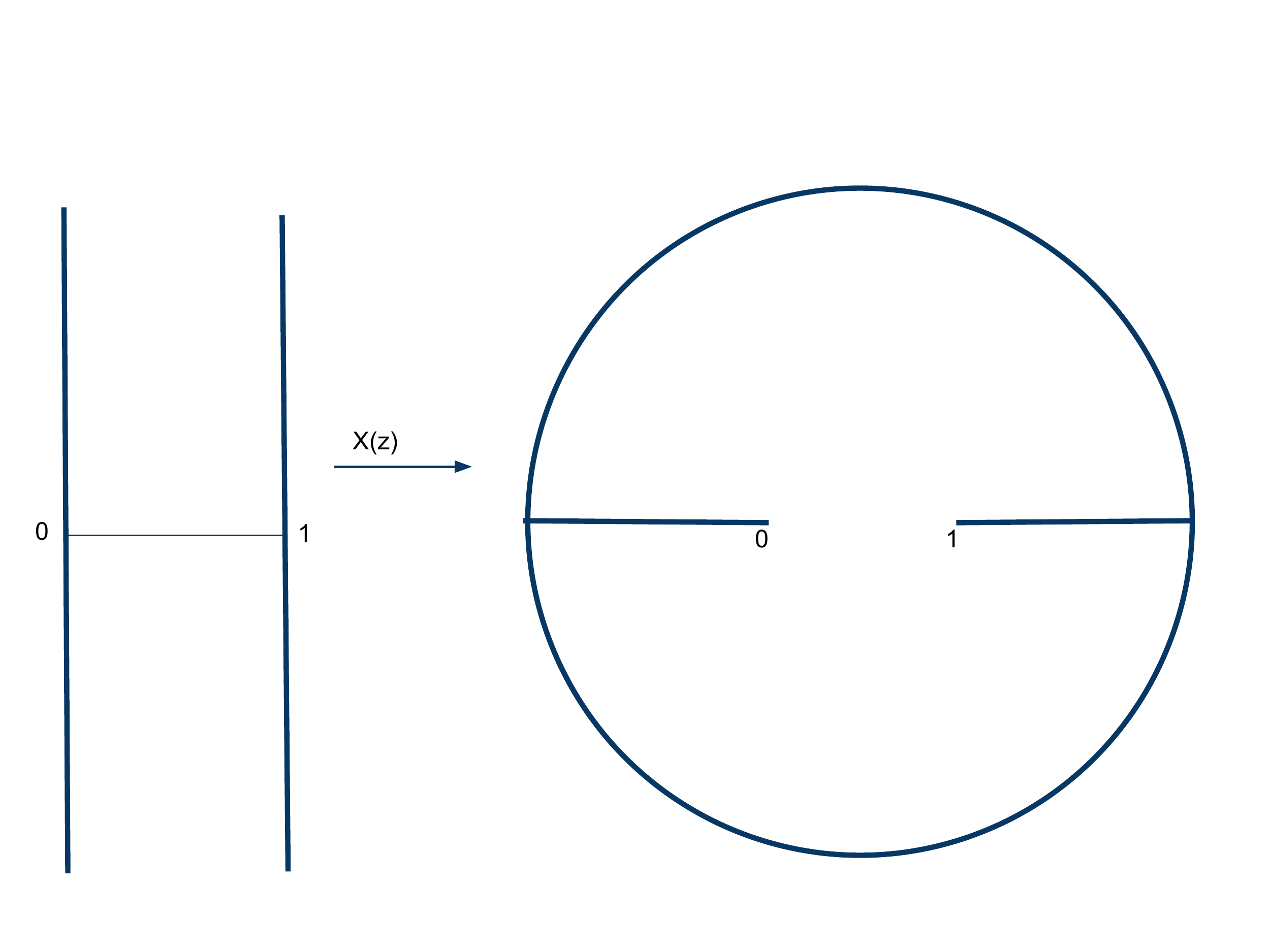}}
\end{center}
\begin{center}
Fig. 5: Belyi map from strip to the complex plane.
\end{center}

For the correlator $\langle \prod_i^n {\rm Tr} M^{2k_i} \rangle$ (with $k=\sum_i k_i$) we have $k$ edges in all the (connected) Feynman graphs that contribute to it. Thus with each of the $k$ strips covering the target space ${\mathbb P}^1$ once, we have a degree $k$ cover. To see how this map is branched, we need to look more closely at the gluing rules for the strips. We observe from the rules given in Sec. 3 that the nontrivial piece of the gluing occurs at the endpoints $(0,1)$ as well as at $\infty$. Let us start with 
$\infty$. Note that the latter is the point where the operators ${\rm Tr} M^{2k_i}$ are inserted. At each such insertion we have $k_i$ strips glued together using the exponential map
\be{}
u(z)=\exp{i\pi z\over k_i}.
\ee 
This is what we had in \eq{vertmap} except that we have substituted the perimeter to be $2k_i$. 
We see that the behaviour near $u=0$ of the map \eq{belstrip} is given by
\be{infbr}
X(u)\propto -u^{-k_i}.
\ee
Thus we have a branching of order $k_i$ at $\infty$ associated to the vertex insertion of 
${\rm Tr} M^{2k_i}$. We have a similar branching at $\infty$ from each of the $n$ insertions leading to a net ramification profile labelled by the cycle $(k_1)(k_2)\cdots (k_n)$. Thus we have (locally) a degree $k$ covering labelled by the above conjugacy class in $S_k$.  Thus given a set of operator insertions in the matrix model, this branching profile is fixed. 

Next consider the points $0$ and $1$. These correspond to the faces of the graph which are being glued up. 
For a face with $m$ sides, we have, on composing with $w(z) \propto z^{2\over m}$, the behaviour near $z=0$
\be{zerobr}
X(w) \propto w^{m}.
\ee
Thus we have a branch point of order $m$. We have a similar branching at $1$ and the net profiles are determined by the particular wick contractions. For every inequivalent contraction we have a branched covering. 

While we have glued together the strips locally, we also have to take into account the following global constraint. This comes from the fact that the two ends of the interval $[0,1]$ are mapped under (\ref{belstrip}) to distinct points (namely $(0,1)$ respectively). Thus if we look at the graph formed from the horizontal edges $[0,1]$ of each of the strips (i.e. the dual to the Feynman graph), the 
alternate vertices must be mapped to distinct points. 
This is the same as demanding that the (dual) graph is bipartite. Equivalently, in terms of the dual of the dual graph (which is the original Feynman graph itself) adjacent faces (across an edge) must be coloured with one of two distinct colours - what is known as bichromatic. Now a genus zero graph in which all vertices have {\it even} valency - as in the planar Feynman graphs for our correlator - is known to be bichromatic. Thus the riemann surface formed from gluing together the strips corresponding to such a graph will admit a Belyi map of 
degree $k$ which is locally given by (\ref{belstrip}).

This is not, in general, possible for graphs of higher genus. The simplest example is to consider is 
the genus one graph in the self contractions for $\langle  {\rm Tr} M^{4} \rangle$. There is only 
one vertex (for the dual graph) and so the above Belyi map is not admissible.\footnote{We thank R. de Mello Koch and S. Ramgoolam for pointing out this example.} But this should not come as too much of a surprise since we expect that the dictionary between single particle string states and gauge invariant correlators becomes more complicated at higher genus due to the mixing between single trace and multi-trace operators. In other words, the correlators $\langle \prod_i^n {\rm Tr} M^{2k_i} \rangle$ will not be natural objects for comparing to an $n$-point string amplitude at higher genus. It is only at the classical (planar) level that we expect this to be given by an $n$-point string scattering amplitude. It will be interesting to work out the correct change of basis and see what the actual Belyi maps are like. 

We note that there is a $2k$ fold cover of the  ${\mathbb P}^1$ via clean Belyi maps. This was, indeed the original construction in \cite{mulpenk1}. These are again given by gluing maps from the strip where the map is now given by 
\be{belstrip2}
\tilde{X}(z)=4X(1-X)=\sin^2{\pi z}. 
\ee
This is now a  {\it two fold} cover of the  ${\mathbb P}^1$ which maps both end points $[0,1]$, of the horizontal edges, to the point $0 \in {\mathbb P}^1$. In addition, the point $z=\half$ is also a simple branchpoint which is mapped to $1 \in {\mathbb P}^1$. When the strips are glued together $(0,1)$ are branchpoints with a behaviour as in (\ref{zerobr}). The behaviour near infinity now reads as
 \be{infbr2}
\tilde{X}(u)\propto -u^{-2k_i}.
\ee
Thus we have a cycle structure at $\infty$ which is $(2k_1)(2k_2)\cdots (2k_n) \in S_{2k}$. The cycle structure at $1$ is of the form $[2^k]$. The cycle structure at zero is now determined in terms of the other two. There is no global constraint and this is a clean Belyi map of the kind that entered into the combinatorial interpretation of \cite{Koch:2010zza}. We see that we can associate a clean Belyi map (locally given by $\tilde{X}$) to every Belyi map (given locally by X). 

We will see, in the next section, how this kind of clean map is reflected in the target space. However, from the point of view of a conventional worldsheet description it is less clear that clean Belyi maps are fundamental. As we will argue in Sec. 6, for the topological string on ${\mathbb P}^1$ it seems more natural to consider maps of degree $k$ rather than $2k$. 

Therefore in the next section we will restrict ourselves to the planar graphs for the correlators $\langle \prod_i^n {\rm Tr} M^{2k_i} \rangle_c$ which admit an interpretation in terms of general Belyi maps of degree $k$. 

\section{Skeletal AdS/CFT}

We now shift to a target space perspective. 
Given that these planar correlators can be rewritten as an explicit sum over Belyi maps of the  
local form (\ref{belstrip}) we can ask whether this gives a sensible AdS/CFT like  interpretation in target space. We will see that there is indeed a natural stringy interpretation to these maps which fits in nicely with many things we have grown familiar to, from AdS/CFT. In a way we see many of the properties of the AdS/CFT mapping in a skeletal way exposing some of the inner workings of the duality. As emphasised in the introduction we arrive at this "X-ray picture" without invoking any string dualities but rather entirely from the structure of  the matrix model. 

Let us first take a closer look at the target space of these Belyi maps. The first thing to notice is that
(using (\ref{belstrip})) we can push forward the quadratic differential on the strip as  
\be{quadiff}
dz^2={dX^2\over \pi^2X(1-X)}= {1\over \pi^2}({dX\over  Y})^2 \equiv \omega^2
\ee
where
$Y^2=X(1-X) =-(X-\half)^2+{1\over 4} ={1\over 4}\tilde{X}$. Here $\tilde{X}$ is given in (\ref{belstrip2}).  
This is the equation for a double cover of the $X$-plane 
${\mathbb P}^1$ and $\omega ={1\over \pi}{dX\over Y}$ 
is the canonical abelian differential (of the first kind) on this riemann surface. 
We recognise this to be the equation for the riemann surface associated with the (complexified eigenvalues of the) Gaussian matrix model. This surface is a double cover of the $X$ plane. The discontinuity 
across its branch cut gives (after a trivial shift and rescaling of $X$)  the Wigner semi-circular law. 
This riemann surface has been associated in the past with the target space for the string dual to the 
matrix model.\footnote{See \cite{Maldacena:2004sn} for a recent discussion - though note that this work considers the non-perturbative (in $g_s$) aspects of the target space geometry (in which the cut disappears). The discussion here, on the other hand, 
is purely perturbative (and at the moment classical/planar).} 

In anycase, the combinatorial fact that the planar correlators can be written in terms of clean Belyi maps of degree $2k$ can be nicely realised. We have, associated to the Feynman diagrams for the correlators, a sum over arbitrary Belyi maps (of degree $k$) 
from the worldsheet to the $X$ plane (given locally by (\ref{belstrip})). We then have the map
$Y^2=X(1-X)$ which converts the general Belyi map into a clean one corresponding to the fact that now the point $X=\half$ has the profile $[2^k]$. 

The explicit Belyi map (\ref{belstrip}) gives a detailed picture of the behaviour of the string in the target space. As we saw in the previous section (using (\ref{infbr})) the branching profile at $\infty$ in the target space 
${\mathbb P}^1_X$ is $(k_1)(k_2)\ldots (k_n)$. This is  the image of the point where the "gauge theory" operators are inserted. In other words,
we can view the operator ${\rm Tr} M^{2k_i}$ creating a branchpoint of order $k_i$. 
The operators are all inserted at $\infty$ which effectively acts as the boundary of the target space 
${\mathbb P}^1_X$ (now visualised as the complex plane plus the point at $\infty$ - See Fig. 5).\footnote{It is perhaps more accurate to view the target space as in the figure, namely, the complex plane with a one dimensional boundary on which the operators are inserted. However the ordering and position of the operators is irrelevant and so effectively the boundary can be viewed as a point and the target space as ${\mathbb P}^1_X$. We thank Shlomo Razamat and Ashoke Sen for discussions on this point.}
It corresponds to the "ultraviolet" of the "spacetime" of the matrix model which in this case 
is just a point. Thus the initial data for scattering amplitudes is specified exactly like in usual AdS/CFT.

On the other hand the interval $[0,1]$ is an "emergent space" for the matrix model corresponding to the support of the eigenvalue density. This is a quantum mechanically emergent space due to the repulsion of eigenvalues. We see that the Belyi maps are nothing other than the worldsheets of individual strings coming in from $\infty$ and interacting on the spatial interval $[0,1]$.  (Note that the spatial section $[0,1]$ on each worldsheet strip is mapped via (\ref{infbr}) to the target space interval $[0,1]$.) In fact, the string interactions occur only at the endpoints $0,1$ of the interval.

Thus once we specify the vertex operators we specify the branching data at $\infty$ and then the corresponding worldsheets interact (split and join) in an arbitrary way but only at the endpoints of the interval $[0,1]$.
This is reminiscent of the Gross-Mende picture for high energy scattering of strings \cite{Gross:1987ar}. 
In fact, the saddle worldsheets that enter into the Gross-Mende analysis have holomorphic and (anti-holomorphic) pieces which have a similar branching at the points of insertion of the external 
vertex operators. 
There is also a similar flavour to a general picture for topological strings proposed by Eynard and Orantin \cite{Eynard:2009qr}.

From the AdS/CFT point of view these worldsheets are the stringy Witten diagrams that replace the point particle propagation and interaction in the bulk that we see in the gravity regime. As has been stressed, this comes naturally 
out of the combinatorial structure of the matrix model. 

\begin{center}
\vskip 2pt
\resizebox{280pt}{200pt}{\includegraphics{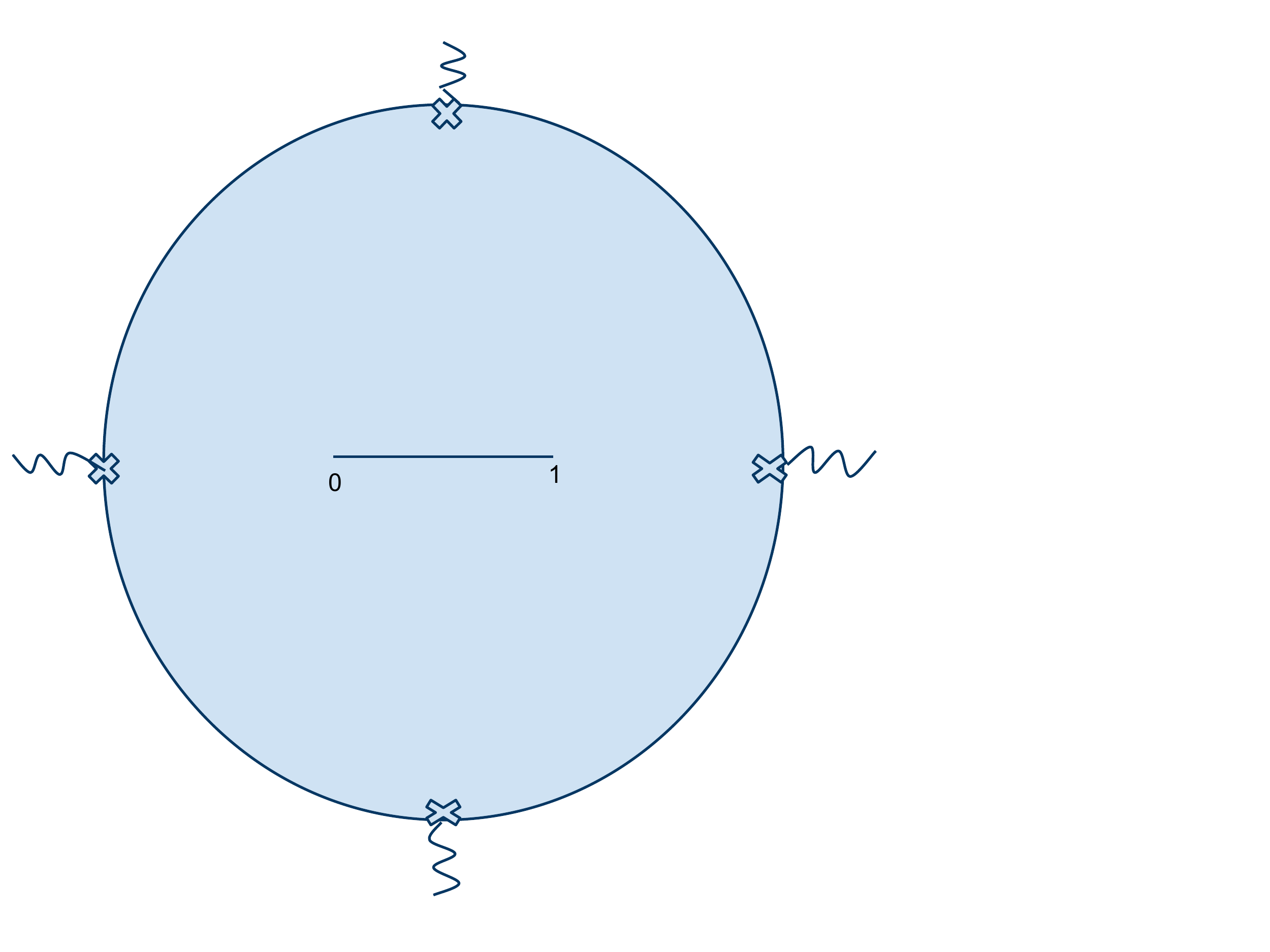}}
\end{center}
\begin{center}
Fig. 6: Matrix Model Operators at $\infty$ create branching of string worldsheets which interact at $0,1$.
\end{center}

One also sees a baby version of holography in this description. For instance, the planar one point function $\langle {\rm Tr} M^{2k}\rangle$ is given by
\be{}
\langle {\rm Tr} M^{2k}\rangle = \int_{cut} \rho(\lambda)\lambda ^{2k} = \oint_{\infty}y(X) X^{2k} dX .
\ee
We can view the first equality as an evaluation on the saddlepoint configuration over the cut in the $X$ plane which maybe viewed as the IR emergent space. The second equality follows from holomorphy and involves the deformation of the contour of integration from around the branch cut to near $\infty$ - the UV of the theory. This latter form is more natural from the conventional 
AdS/CFT dictionary  where we evaluate gauge theory correlation functions (at large $N$) by a saddle point evaluation which reduces to a boundary term at infinity.

\section{Relation to the A-Model Topological String Theory on ${\mathbb P}^1$}

One might wonder what the stringy picture, that we have sketched above, has to do with conventional string theories. In other words, what is the conventional worldsheet string theory description which manifests all the above properties. At first sight, there are definitely some unusual looking features to the sum over worldsheets interpretation to the gaussian correlators. It may not be obvious to the reader that these features can indeed be realised in a standard string theory.  

The first unusual feature is that we have a discrete sum over worldsheet riemann surfaces and not an integral over the entire moduli space. This is because, as mentioned earlier, Belyi maps are admissible only for those riemann surfaces which are artithmetic (having a defining equation over the field $\bar{\mathbb Q}$). However, this feature of the worldsheet theory getting contributions from discrete points on the moduli space of riemann surfaces is not very new. 
In the context of topological string theories this has been known for a while though it is perhaps not very familiar.\footnote{Parenthetically, we also note that the prescription \cite{Gopakumar:2005fx} to map proper time correlation functions to integrands on moduli space generically exhibits 
for free field theories, a milder kind of localisation on lower codimension subspaces \cite{Aharony:2006th},. See \cite{Aharony:2007fs, David:2008iz} for further illustrations of this as well as discussion.}

For instance, consider the A-model string theory with a target space which is a $T^2$. In addition, specialise to the case where the worldsheet is of genus one. Then we know that the worldsheet path integral localises on holomorphic maps from $T^2_{WS}$ to $T^2_{TS}$. Such holomorphic maps (for a given degree $k$) exist only at discrete points in the fundamental key hole region of the moduli space ${\cal M}_{1,0}$. This is explicitly worked out in \cite{Bershadsky:1993ta} and the discrete set of points are given by 
\begin{equation}
\tau_{WS}={m\pm r\tau_{TS}\over k} \qquad (r \geq 1; m=0,1, \ldots (k-1) ). 
\end{equation}
One can see from here that the phenomenon of localisation on the moduli space is not uncommon and in some sense generic to the A-model topological string where one is restricted to holomorphic maps.

Closer to the context of the present paper, it was shown in \cite{Verlinde:1990ku} that amplitudes in minimal topological string theories (including topological gravity) are also localised on moduli space. In particular, a gauge choice of worldsheet metric was made such that all the contributions to amplitudes come from delta function contact terms at the locations of the vertex operator insertions. Something similar appears to happen in the Strebel worldsheet gauge such that one localises to maps involving only three branchpoints. In fact, in the Strebel gauge one has a locally flat metric except at the insertions of the vertex operators (poles) {\it and} interaction vertices (zeroes). Compare with the comments below
Eq. (4.13) of  \cite{Verlinde:1990ku}.

The second unusual looking feature is that the contribution to a particular matrix model correlator 
$\langle {\rm Tr} M^{2k_1}\ldots  {\rm Tr}M^{2k_n}\rangle_c$ comes {\it only} from holomorphic maps 
of a fixed degree, namely, $k=\sum_ik_i$. In familiar A-model topological string theories with  target Calabi-Yau manifolds, one typically has contributions to amplitudes 
from maps of arbitrary (positive) degree. 
However, as we will now review, this is a feature specific to target spaces with $c_1(M)=0$. In fact, we will show that a conventional A-model topological string theory on ${\mathbb P}^1$ has {\it exactly the right features to precisely reproduce the above behaviour}. 

Recall that the A-model theory can be defined on any Kahler target space. The physical observables consist of the the identity or puncture operator $P$ and its gravitational descendants $\sigma_n(P)$ together with that corresponding to other topological primaries $Q_{\alpha}$ 
(such as the Kahler class $Q$)
and its descendants denoted by $\sigma_n(Q_{\alpha})$. 

There is an important 
selection rule for such A-model correlators which is essentially a ghost number conservation law 
\cite{Witten:1989ig}.
In general, the genus $g$ correlator $\langle \prod_{i=1}^n\sigma_{m_i}(Q_{\alpha_i}) \rangle_g$ gets contributions only from holomorphic maps of degree $d$ which obey (see for e.g. Eq.(2.2) of \cite{Eguchi:1996tg})
\be{selrule}
\sum_{i=1}^n(m_i+q_{\alpha_i}-1)=(3-D)(g-1)+c_1(M)d.
\ee
Here $D$ is the complex dimension of the target space $M$ and $c_1(M)$ its first chern number. While $q_{\alpha_i}$ is the complex dimension of the class ($U(1)$ charge) corresponding to $Q_{\alpha_i}$. 
Thus for a target space where $c_1(M)=0$, the term involving the degree drops out. (In the case of Calabi-Yau three folds, the first term on the RHS also vanishes and hence we do not have any nontrivial correlators involving descendants.)

In the case of target ${\mathbb P}^1$, we have $D=1$, $c_1=2$ and $q=0,1$ for $P,Q$ respectively. 
Thus for a correlator $\langle \sigma_{2k_1-1}(Q)\sigma_{2k_2-1}(Q)\sigma_{2k_3}(Q)\ldots 
\sigma_{2k_n}(Q)\rangle_0$, the LHS of (\ref{selrule}) is $-2+2\sum k_i$.\footnote{The factor of $(-2)$ comes from having taken two of the descendant operators to have indices $(2k_i-1)$.}  
The RHS is given 
(for $g=0$) by $-2+2d$. Thus we conclude that such a correlator is nonvanishing only when 
\be{selrule2}
d=\sum_{i=1}^nk_i=k.
\ee

But this is precisely the constraint that our worldsheet picture for the Feynman graphs gave us! We thus arrive at a rough identification $\sigma_{2k}(Q) \leftrightarrow {\rm}Tr M^{2k}$. Or rather,\footnote{
We have treated operators 1 and 2 in a special way.
One obviously should sum over the various different permutations in the indices. Presumably there is a better way to write this in terms of $\prod_i \sigma_{2k_i}(Q)$ with extra insertions of  puncture operators P.}
\be{gausstopcorr1}
\langle {\rm Tr} M^{2k_1}\ldots  {\rm Tr}M^{2k_n}\rangle_c \sim \langle \sigma_{2k_1-1}(Q)\sigma_{2k_2-1}(Q)\sigma_{2k_3}(Q) \ldots \sigma_{2k_n}(Q)\rangle_0.
\ee
It is a somewhat nontrivial match with the sum over Belyi maps picture that such a consistent identification is even possible which gives exactly the same degree.  
As mentioned in Sec. 4, the single trace correlators are the natural gauge invariant operators dual to single string vertex operators at the classical level. Hence we will restrict ourselves, in this work, to the genus zero case.  
We will thus take the 
right hand side to be the genus zero topological correlator. We will furthermore evaluate it 
at the origin of the phase space of the various couplings, i.e. without any background turned on.

We offer some preliminary evidence in favour of the identification (\ref{gausstopcorr1}) though more work is needed to make this precise.

First, consider the simplest case of a planar one point function $\langle {\rm Tr} M^{2k}\rangle_c$ in the matrix model. The answer is the well known Catalan number
\be{mmonept}
\langle {\rm Tr} M^{2k}\rangle_c = {2k!\over k!(k+1)!} \equiv C_k.
\ee
Since the one point function is special, we will identify this with $\langle \sigma_{2k-1}(Q) P \rangle_0$. We find 
\be{toponept}
\langle \sigma_{2k-1}(Q)P \rangle_0 = (2k-1)  \langle \sigma_{2k-2}(Q) \rangle_0  = (2k-1)\times {(2k-2)!\over k!^2}.
\ee
Here we have used the puncture equation \cite{Dijkgraaf:1990nc} and finally used the already calculated value of
$\langle  \sigma_{2k-2}(Q) \rangle_0$ \cite{Eguchi:1995er}.

We see that 
\be{}
{1\over 2k} \langle {\rm Tr} M^{2k}\rangle_c = {1\over k+1} \langle  \sigma_{2k-1}(Q) P \rangle_0.
\ee

A somewhat more nontrivial check is that of the connected planar two point function 
$\langle {\rm Tr} M^{2k_1}{\rm Tr} M^{2k_2}\rangle_c$. Note that the connected piece is down by two powers of $N$ from the disconnected contribution which follows from large $N$ factorisation. 
This piece is less easy to calculate but the answer can be extracted from results in the matrix model literature. This is done in Appendix A where we find 
\be{}
\langle {\rm Tr} M^{2k_1}{\rm Tr} M^{2k_2}\rangle_c = C_{k_1}C_{k_2}\times {k_1(k_1+1)k_2(k_2+1)\over (k_1+k_2)}.
\ee
Here, $C_k$ is the Catalan number defined in (\ref{mmonept}). 

We can also evaluate $\langle \sigma_{2k_1-1}(Q) \sigma_{2k_2-1}(Q) \rangle_0$ using topological recursion relations (see appendix B for details)
\be{}
\langle  \sigma_{2k_1-1}(Q) \sigma_{2k_2-1}(Q) \rangle_0 = C_{k_1}C_{k_2}\times {(k_1+1)(k_2+1)\over 4(k_1+k_2)}.
\ee
Therefore,
\be{}
{1\over (2k_1)(2k_2)} \langle {\rm Tr} M^{2k_1}{\rm Tr} M^{2k_2}\rangle_c
= \langle  \sigma_{2k_1-1}(Q) \sigma_{2k_2-1}(Q) \rangle_0
\ee

Thus once, again we find that the two answers are simply related to each other. Note that the answer (of both the matrix model and topological string theory) does not have a factorised dependence on 
$k_1, k_2$ and hence we find this matching somewhat nontrivial. 

Eguchi and Yang \cite{Eguchi:1994in, Eguchi:1995er} have proposed a matrix integral which reproduces the amplitudes of the topological string theory on ${\mathbb P}^1$. On a subset of the couplings (mostly gravitational descendants of $Q$ turned on) this model reads as 
\bea{egyang}
Z[\tilde{M}] &=& \int[d\tilde{M}]e^{-N{\rm Tr}S(\tilde{M})} \cr
S(\tilde{M}) &=& 2(t_{1,P}-1)\tilde{M}(\ln \tilde{M}-1)+ \sum_{n=1}{t_{n-1,Q}\over n}\tilde{M}^n.
\eea
Here $t_{n,Q}$ is the coupling to the descendant $\sigma_n(Q)$ (and similarly for $P$).

Thus the Eguchi-Yang model 
appears to be very similar to our case, but there are some important differences as well. 
Firstly, one needs to get rid of the logarithmic term by setting $t_{1,P}=1$. This is somewhat singular. Secondly, we need a quadratic action to evaluate the Gaussian correlators. In (\ref{egyang}) this corresponds to a background of $t_{1,Q}$. Thirdly, this model has an identification which is
$\sigma_{2k-1}(Q)\leftrightarrow {\rm Tr}\tilde{M}^{2k}$. It would be good to understand explicitly the relation between this model (which is known to capture the Toda integrability of the string background) and Gaussian correlators.  We also note that topological strings on ${\mathbb P}^1$ have also made an appearance in relation to the BPS sector of ${\cal N}=2$ $U(1)$ (noncommutative) 
gauge theory on $R^4$, which is again closely related to matrix models \cite{Losev:2003py}.  

It is also amusing that the Gaussian matrix model with a modified (unitary) measure is equivalent to the topological Chern-Simons theory on $S^3$ \cite{Marino:2002fk}. Through the open-closed string duality, 
\cite{Gopakumar:1998ki} 
this modified Gaussian model therefore describes the topological closed string on the local ${\mathbb P}^1$ geometry \cite{Aganagic:2002wv}. It is as if the additional normal directions to the ${\mathbb P}^1$
in the resolved conifold change the measure to the unitary one. See \cite{Marino:2004eq} for a review of the connection between matrix models and topological string theories. 

We have begun here to make a concrete connection between the Feynman graph worldsheets and 
the A-model. It is important in this context to understand from the conventional A-model point of view, how the Strebel gauge pick out maps with precisely three branchpoints. In this context, we note that Kontsevich has used techniques of localisation to study topological strings on 
toric spaces including ${\mathbb P}^1$. His answer is expressed in terms of graphs which have many similarities with that of the Gaussian matrix model. It will be nice to understand whether this is the relevant mechanism at work here.\footnote{We thank Tom Brown for bringing \cite{Kontsevich:1994na} 
to our attention and for useful discussions on these matters.}

\section{Discussion}

Let us summarise what we have learnt as well as what needs to be understood better. 

Firstly, we have a natural worldsheet realisation of the combinatorial fact that Gaussian correlators can be associated to Belyi maps. This worldsheet realisation is in 
the unusual Strebel gauge where all the curvature is concentrated at either the vertices (of the Feynman graph) or the centers of faces (equivalently at the insertions  vertex operators as well as interactions).   In particular, it was crucial to interpret the Feynman diagrams {\it a la} Razamat where the strebel length took integer values. This fits in exactly with the discrete triangulation approach but with a different interpretation, namely, of discrete contributions from points in moduli space which correspond to the Belyi maps. 

Secondly, the target spacetime that emerges from this point of view is nothing but the natural riemann surface associated to the (eigenvalues of the) Matrix model. This is again something that has been advocated before and is fitting from several points of view. Specifically, this riemann surface is known to  capture all the gauge invariant information in the matrix model (for all genus) and it is thus apt that it be the target space geometry for the dual string. 

Thirdly, we find a AdS/CFT like picture in this target spacetime for the correlators 
of the gauge theory. This picture is one in which external strings create a specific branching as they come in from infinity (UV of the matrix model). They then split and join at specific points (in the IR)  in a very natural stringy generalisation of the Witten diagrams in AdS that appear in the supergravity limit. 
The basic picture is also reminiscent of the Gross-Mende description which is believed in some sense to capture some universal features of string interactions. 

Fourthly, we found many of the unusual features of the worldsheet description such as the localisation on moduli space as well as contributions from fixed degree maps can be realised in a standard A-model topological string theory on ${\mathbb P}^1$. We also found evidence for the observables in this string theory capturing matrix model correlators. 

Thus the Gaussian matrix model puts together many pieces of intuition about gauge-string duality (both in general as well as for the particular case of matrix models) in a nice geometric setting. 

But there are many issues that need to be explored in detail. Firstly, it is important to understand the Strebel worldsheet gauge better even if in the limited context of standard topological string theory. Does it generically lead to a fragmentation of the worldsheet into strips which have an interpretation as matrix model Feynman diagrams? Secondly, it would be good to understand why, in the 
theory on ${\mathbb P}^1$, this gauge leads to only three branchpoints. It is likely that the answer will involve the localisation picture of Kontsevich \cite{Kontsevich:1994na}. Thirdly, the connection of observables and correlators of the topological string on ${\mathbb P}^1$ with that of the matrix model needs to verified and refined further. Is this nothing other than the Eguchi-Yang Matrix model suitably interpreted? 
Fourthly, many of our considerations were restricted to the planar/classical level since the basis of single trace operators is natural only at that level. How exactly do we go beyond genus zero and develop a dictionary? Note that the connection to clean Belyi maps exists for all genus. 
Finally, It would  be very nice to understand better the mathematical significance of the appearance of Belyi maps which play a very important role in understanding the absolute Galois group. Can we use matrix models to make nontrivial statements about the absolute Galois group? 
See some explorations in this direction in \cite{Koch:2010zza}.

\bigskip

{\bf Acknowledgements:} We would like to thank 
R.~de Mello Koch and S.~Ramgoolam for stimulating discussions 
at the JoBurg Workshop (Apr. 2010) which led to this project. 
We would also like to thank
T.~Brown, A.~Dabholkar, 
J.~R.~David, B. ~Ezhuthachan, M.~R.~Gaberdiel, N.~Iizuka, G.~Jones, 
G.~Krishna, D.~Kutasov, 
M.~Marino, S.~Minwalla, S.~Mukhi, N.~Orantin, K.~Papadodimas,
S.~Pasquetti, B.~Pioline, M.~Rangamani, S.~Razamat, A.~Sen,
S.~Wadia and especially  E.~Verlinde for useful 
discussions and correspondence. Thanks are also due to T. ~Brown, S.~Razamat and A.~Sen for their helpful comments on the manuscript.  
We would like to express thanks to the 
International Centre for Theoretical Sciences (ICTS) and CHEP (IISc, Bangalore), CTS (ETH Zurich), University of Amsterdam, CERN (Geneva) and LPTHE (Paris) for the hospitality during various stages of this work. We also thank the organisers of the mini-String workshop (Paris), LMS (Durham) and Ascona (Monte Verita) meetings for the opportunity to present many of the results of this paper. 
This work was partly supported by a 
Swarnajayanthi Fellowship of the Dept.\ of Science and Technology, Govt.\ of India 
and as always by the support for basic science by the Indian people.

\section*{Appendix} 
\appendix

\section{Two Point functions in the Gaussian Matrix Model}

We would like to evaluate 
\be{}
C_{2k_1,2k_2} \equiv \langle {\rm Tr} M^{2k_1}{\rm Tr} M^{2k_2}\rangle_c.
\ee 
There exist closed form expressions for generating functions of $C_{2k_1,2k_2}$ from which these numbers can be extracted. See, for instance, \cite{Morozov:2009uy, Morozov:2010uu}. We will find 
it useful to use the particular form employed in \cite{Akemann:2001st}. 
Define
\be{}
W_0(x)=\langle {\rm Tr} (e^{xM})\rangle_c; \qquad W_0(x,y)=\langle {\rm Tr} (e^{xM}){\rm Tr} (e^{yM})\rangle_c
\ee
where the subscript $0$ refers to the planar contribution. Explicit expressions for $W_0(x), W_0(x,y)$
are given by  \cite{Akemann:2001st} to be\footnote{To compare with the expressions in 
\cite{Akemann:2001st}, note that 
we have fixed their parameter $a=2$ which gives the correctly normalised expressions for us.} 
\be{}
W_0(x)={1\over x}I_1(2x); \qquad  W_0(x,y)={xy\over x+y}(I_0(2x)I_1(2y)+I_1(2x)I_0(2y)).
\ee
Here $I_{\nu}(z)$ is the modified Bessel function which has the series expansion
\be{}
I_{\nu}(z)=\sum_{k=0}^{\infty}{1\over k! \Gamma(\nu+k+1)}({z\over 2})^{\nu+2k}.
\ee

In particular, as a check on normalisations etc., we see that 
\be{}
W_0(x)={1\over x}I_1(2x)=\sum_{k=0}^{\infty}{1\over k!(k+1)!}x^{2k}=\sum_{k=0}^{\infty}{C_k\over (2k)!}
x^{2k}
\ee
implying that $\langle {\rm Tr} M^{2k}\rangle_c= C_k ={(2k)!\over k!(k+1)!}$ as expected.

We can now go on to look at $W_0(x,y)$. 
\bea{}
(x+y)W_0(x,y) &=& xy(I_0(2x)I_1(2y)+I_1(2x)I_0(2y)) \cr
&=& \sum_{k_1,k_2=0}^{\infty}C_{k_1}C_{k_2}{x^{2k_1+1}\over (2k_1)!}{y^{2k_2+1}\over (2k_2)!}[(k_2+1)x+(k_1+1)y] 
\eea
On the other hand, by definition,
\bea{}
(x+y)W_0(x,y) = (x+y) \sum_{k_1,k_2=0}^{\infty} [&&C_{2k_1,2k_2}{x^{2k_1}\over (2k_1)!}{y^{2k_2}\over (2k_2)!} \cr
&& + C_{2k_1+1,2k_2+1}{x^{2k_1+1}\over (2k_1+1)!}{y^{2k_2+1}\over (2k_2+1)!} ] 
\eea

Comparing powers, and subtracting the contributions of odd moments, gives us a recursion relation for the even moments
\be{}
C_{k_1}C_{k_2}(k_1-k_2)=C_{2k_1, 2k_1+2}{1\over (2k_1+2)(2k_2+1)}-C_{2k_1+2, 2k_1}{1\over (2k_1+2)(2k_1+1)}.
\ee
It is easy to check that this recursion relation is solved by 
\be{}
C_{2k_1,2k_2}=  C_{k_1}C_{k_2}\times {k_1(k_1+1)k_2(k_2+1)\over (k_1+k_2)}.
\ee

For small values of $k_1,k_2$ one can check that this expression gives the correct combinatoric factors from the graphical expansion. 
(See for e.g. the expressions in \cite{Alexandrov:2003pj} - Eq.(III.1.35)).
Also one can check that  $\langle {\rm Tr} M^{2}{\rm Tr} M^{2k}\rangle_c =(2k)\times C_k$ which agrees with the above expression.

\section{Computation of $\langle \sigma_{2k_1-1}(Q) \sigma_{2k_2-1}(Q) \rangle_0$}

We use various recursion relations for correlators in the topological string to evaluate 
$\langle \sigma_{2k_1-1}(Q) \sigma_{2k_2-1}(Q) \rangle_0$. To evaluate one, two and three point functions requires a little care.  Fortunately, recursion relations for the relevant cases have been worked out in the literature.

In particular, we use the recursion relation ("Fundamental Recursion Relation") employed by Eguchi, Hori and Xiong 
(see \cite{Eguchi:1996tg} - Eq. (2.13)).
On specialising to ${\mathbb P}^1$, this read as 
\be{}
\langle \sigma_{m}(Q) \sigma_{n}(Q) \rangle_0 ={nm \over n+m+2}{\cal M}^{\rho\sigma}
\langle \sigma_{m-1}(Q) {\cal O_{\rho}} \rangle_0 \langle  \sigma_{n-1}(Q)  {\cal O_{\sigma}} \rangle_0.  
\ee
Here ${\cal O_{\rho}}$ are the topological primaries which in our case are just $P,Q$. The matrix 
${\cal M}^{\rho\sigma}$ has non-zero entries (at the origin of couplings $t_{n,P}=t_{n,Q}=0$) 
${\cal M}^{PP}={\cal M}^{QQ}=2$
(see Eq. (3.32) of \cite{Eguchi:1996tg}).

Using the fact that 
$\langle \sigma_{2k-2}(Q) P\rangle_0 =(2k-2) \langle \sigma_{2k-3}(Q) \rangle_0 =0$ (since the puncture equation can be shown to hold for this two point function and the last correlator vanishes from the selection rule.) we have the only non-vanishing contribution
\bea{}
\langle \sigma_{2k_1-1}(Q) \sigma_{2k_2-1}(Q) \rangle_0&=&{(2k_1-1)(2k_2-1)\over 2(k_1+k_2)}
{\cal M}^{QQ}\langle \sigma_{2k_1-2}(Q) Q \rangle_0 \langle \sigma_{2k_2-2}(Q) Q \rangle_0 \cr
&=&{ k_1k_2 (2k_1-1)(2k_2-1)\over (k_1+k_2)}\langle \sigma_{2k_1-2}(Q) \rangle_0\langle  \sigma_{2k_2-2}(Q)\rangle_0 \cr
&=& {(k_1+1)(k_2+1)\over 4(k_1+k_2)}\times C_{k_1}C_{k_2}
\eea
In the second line we have used the relation $\langle \sigma_{2k}(Q) Q \rangle_0 =(k+1)\langle \sigma_{2k}(Q) \rangle_0$ \cite{Eguchi:1996tg}. Finally, we also used the result for the one point function
$\langle \sigma_{2k}(Q) \rangle_0 = {(2k)! \over (k+1)!(k+1)!}$ \cite{Eguchi:1995er}.

\bibliographystyle{JHEP}

\end{document}